\documentclass[twocolumn,preprintnumbers,notitlepage,superscriptaddress,amsmath,10pt,aps,longbibliography]{revtex4-1}

\usepackage{graphicx}
\usepackage{xcolor}
\usepackage{hyperref}
\usepackage{subcaption}
\usepackage{float}
\usepackage{graphicx}
\usepackage{type1cm}
\usepackage{eso-pic}
\usepackage{color}
\usepackage{amsmath}
\usepackage{amssymb}
\usepackage{braket}
\usepackage{gensymb}
\usepackage{bbold}
\usepackage{amsthm}

\usepackage{threeparttable, caption}
\theoremstyle{definition}
\newtheorem{definition}{Protocol}

\begin{document}

\title{Quantum Rabin Oblivious Transfer Using Two Pure States}

\author{Lara Stroh}
\affiliation{SUPA, Institute of Photonics and Quantum Sciences, School of Engineering and Physical Sciences, Heriot-Watt University, Edinburgh EH14 4AS, United Kingdom}

\author{James T. Peat}
\affiliation{SUPA, Institute of Photonics and Quantum Sciences, School of Engineering and Physical Sciences, Heriot-Watt University, Edinburgh EH14 4AS, United Kingdom}

\author{Mats Kroneberg}
\affiliation{SUPA, Institute of Photonics and Quantum Sciences, School of Engineering and Physical Sciences, Heriot-Watt University, Edinburgh EH14 4AS, United Kingdom}

\author{Ittoop V. Puthoor}
\affiliation{SUPA, Institute of Photonics and Quantum Sciences, School of Engineering and Physical Sciences, Heriot-Watt University, Edinburgh EH14 4AS, United Kingdom}
\affiliation{School of Computing, Newcastle University, Newcastle upon Tyne NE4 5TG, United Kingdom}

\author{Erika Andersson}
\email{E.Andersson@hw.ac.uk}
\affiliation{SUPA, Institute of Photonics and Quantum Sciences, School of Engineering and Physical Sciences, Heriot-Watt University, Edinburgh EH14 4AS, United Kingdom}

\begin{abstract}
Oblivious transfer between two untrusting parties is an important primitive in cryptography. There are different variants of oblivious transfer. In Rabin oblivious transfer, the sender Alice holds a bit, and the receiver Bob either obtains the bit, or obtains no information with probability $p_?$. Alice should not know whether or not Bob obtained the bit. 
We examine a quantum Rabin oblivious transfer (OT) protocol that uses two pure states. Investigating different cheating scenarios for the sender and for the receiver, we determine optimal cheating probabilities in each case. Comparing the quantum Rabin oblivious transfer protocol to classical Rabin oblivious transfer protocols, we show that the quantum protocol outperforms classical protocols which do not use a third party, for some values of $p_?$. We find that quantum Rabin OT protocols that use mixed states can outperform quantum Rabin OT protocols that use pure states for some values of $p_?$.
\end{abstract}

\maketitle

\section{Introduction}

Apart from quantum key distribution, other cryptographic primitives have been studied in the quantum setting. One of these is oblivious transfer (OT), where, unlike in (quantum) key distribution, sender and receiver do not trust each other.
OT was initially introduced informally by Wiesner in the context of ``conjugate coding", where a sender encodes two messages, but only one of these messages can be received~\cite{Wiesner83}. Even \textit{et al.}~\cite{Even85} later defined 1-out-of-2 oblivious transfer, where a sender sends two bits to a receiver, who learns only one of them, while the sender does not learn which bit is received.

Other variants of oblivious transfer exist, such as 1-out-of-$n$ OT~\cite{Brassard86}, XOR OT~\cite{Brassard03}, generalized OT~\cite{Brassard03}, and Rabin OT~\cite{Rabin05}. In Rabin oblivious transfer, which is also known as all-or-nothing oblivious transfer, a sender sends a bit to a receiver, who either learns the bit value or learns nothing at all. The sender remains ignorant of whether the receiver obtained the bit value or not.
Cr{\'e}peau~\cite{Crepeau88} showed that Rabin OT and 1-out-of-2 OT are classically equivalent, in the sense that if one of them can be implemented, then this can be used to realise the other one. In fact, all of the above variants have been shown to be equivalent to 1-out-of-2 OT in the classical setting~\cite{Brassard86, Brassard03}. This equivalence may or may not hold for quantum oblivious transfer.

The significance of OT stems from the fact that it can be used to implement any two-party computation~\cite{Kilian88}. Unfortunately, information-theoretically secure (perfect) OT is impossible both in the classical and quantum settings. In classical 1-out-of-2 oblivious transfer, if one of the parties cannot cheat any better than with a random guess, then the other party can cheat with probability equal to 1 without being detected.
Mayers~\cite{Mayers97} and Lo~\cite{Lo97} proved that one-sided quantum two-party computation is impossible with information-theoretic security. Since 1-out-of-2 OT is universal for two-party computation, it follows that (perfect) quantum 1-out-of-2 OT with information-theoretic security must also be impossible.

By introducing realistic restrictions on the communicating parties, information-theoretically secure perfect quantum OT does become possible. For instance, Damg\aa rd \textit{et al.} introduced the bounded quantum storage model and proposed a quantum Rabin OT protocol that is secure with this assumption~\cite{Damgard08}. Another approach involves spacetime-constrained quantum oblivious transfer, which is unconditionally secure in a relativistic quantum setting~\cite{Garcia16, Garcia18, Garcia19}.
Moreover,  even without restrictions on the communicating parties, cheating probabilities can still be limited without violating the no-go results in~\cite{Mayers97} and~\cite{Lo97}. An active area of research focuses on the investigation of what the smallest possible cheating probabilities are for quantum oblivious transfer~\cite{Chailloux13, Chailloux16, Amiri21, Stroh23}.

Quantum Rabin oblivious transfer has been less investigated than other variants of quantum oblivious transfer.
Two quantum protocols for Rabin OT are presented by Bansal and Sikora~\cite{Bansal23}. These are based on switching between two ``bad" such protocols, to create an overall better Rabin oblivious transfer protocol. The security analysis for one of these stochastic switching protocols uses a different definition for a dishonest sender than the one we use, but the security analysis for the other protocol uses the same definition as we do. It can therefore be directly compared to the protocol investigated in this paper. Although the underlying ``bad" protocols achieve the same cheating probabilities as the basic protocol here does, we can obtain lower cheating probabilities for both sender and receiver by including some testing by the receiver than they can obtain by applying the stochastic switching.
Furthermore, a quantum Rabin OT protocol is proposed in~\cite{He06} and is even claimed to be secure. However, it appears that the considered cheating strategies may not be completely general and it can indeed be shown that there are successful cheating strategies for a dishonest party~\cite{Peat24}.

Because of these aspects, it is of interest to investigate quantum Rabin oblivious transfer protocols. The simplest possible quantum Rabin OT protocol is arguably for the sender to encode a bit value in two pure non-orthogonal states. The receiver makes an unambiguous measurement to either perfectly learn the bit value, or fail to learn anything at all~\cite{Cheong10}.
At first this may not seem very promising, since the sender can trivially cheat perfectly by making the receiver always receive a bit value (albeit at the expense of the sender not knowing what bit value the receiver thinks they have received). However, further investigation is needed, since, if the protocol is implemented many times and the receiver monitors how often they receive or do not receive the bit value, they can limit how often the sender can use the perfect cheating strategy. It is also of interest to have a full analysis of the quantum Rabin OT protocol using two pure states, in order to obtain the achievable cheating probabilities and have a baseline for how well Rabin OT with pure states can perform.

In this paper, we evaluate the performance of the quantum Rabin oblivious transfer protocol proposed in~\cite{Cheong10} more fully, and compare it to classical Rabin OT protocols. 
The protocol is described in Section \ref{Protocl}. In Section \ref{Bob cheats}, we consider a cheating receiver, and in Section \ref{alice cheats} a cheating sender. We consider three different cheating strategies for the sender, depending on whether and how the receiver tests the states they receive.
In Section \ref{sec: Comparison}, we evaluate the performance of the quantum Rabin OT protocol. We briefly compare it  to the stochastic switching quantum Rabin OT protocol by Bansal and Sikora~\cite{Bansal23}. The main focus of this section is however the comparison to classical Rabin OT protocols. We show that the quantum advantage in terms of lower cheating probabilities is either small or nonexistent. The comparison with classical protocols suggests that mixed states might help to improve security of quantum Rabin oblivious transfer protocols. By expressing the classical protocol as a quantum protocol using two mixed states, we find that this is certainly sometimes the case.

\section{The Protocol} \label{Protocl}

In Rabin oblivious transfer, a sender Alice has one bit, and sends it to a receiver Bob. Bob should either receive the sent bit or receive nothing. Ideally, Alice should have no information about whether Bob received the bit or not, and Bob should not have any information about the bit value when he did not receive it.
In the quantum Rabin OT protocol we consider~\cite{Cheong10}, Alice encodes her bit value $x \in \{0, 1\}$ using one of the two pure states $\ket{\psi_{x}}$. The overlap between the states can, without loss of generality, be considered to be real and equal to $\braket{\psi_0|\psi_1}=\cos(2\theta)$, where $0 \degree \leq \theta \leq 45 \degree$. The protocol is performed as follows.

\begin{enumerate}
\item Alice randomly chooses a bit value $x$, with equal probability for $x=0$ and $x=1$. She prepares and sends the state $\ket{\psi_x}$ to Bob, where
\begin{align} \label{Honest states}
\ket{\psi_0} = \cos\, \theta \ket{0} + \sin\, \theta \ket{1} , \nonumber \\
\ket{\psi_1} = \cos\, \theta \ket{0} - \sin\, \theta \ket{1} .
\end{align}
\item Bob performs an unambiguous discrimination measurement~\cite{Ivanovic87, Dieks88, Peres88} on the received state. His measurement operators are
\begin{align} \label{Eq: Bob Measurement Operators}
\Pi_0 &= \dfrac{1}{2\, \cos^2\, \theta} \ket{\overline{\psi_1}} \bra{\overline{\psi_1}} , \nonumber \\[0.2em]
\Pi_1 &= \dfrac{1}{2\, \cos^2\, \theta} \ket{\overline{\psi_0}} \bra{\overline{\psi_0}} , \nonumber \\[0.3em]
\Pi_? &= (1-\text{tan}^2\, \theta) \ket{0}\bra{0} ,
\end{align}
corresponding to a bit value of 0 and 1, and not receiving the bit, respectively. 
\end{enumerate}
Bob learns the bit value with probability $1-p_{?}= 1-\cos(2\theta) = 2\sin^2\theta$ and  obtains an inconclusive result with probability $p_{?}= \cos(2 \theta )$~\cite{Ivanovic87, Dieks88, Peres88}.
The states
\begin{align}
\ket{\overline{\psi_0}} &= \sin\, \theta \ket{0} - \cos\, \theta \ket{1} , \nonumber \\
\ket{\overline{\psi_1}} &= \sin\, \theta \ket{0} + \cos\, \theta \ket{1} 
\end{align}
are orthogonal to Alice's states in Eq. \eqref{Honest states}, which guarantees that Bob will never obtain bit value 0 when Alice sent $|\psi_1\rangle$, and vice versa. 
In a Rabin OT protocol, Bob should usually receive the bit value with probability $1/2$, which corresponds to $\theta = 30 \degree$. Other values for $p_?$ are, however, possible. If quantum Rabin OT protocols do give an advantage over corresponding classical protocols, it is not immediately clear for what value of $p_?$ this advantage would be the greatest.

\section{Dishonest Bob} \label{Bob cheats}

Let us assume that Bob is dishonest and Alice is honest (we do not care about a dishonest party being cheated against, and if both parties are dishonest, both can deviate from the protocol). Cheating by Bob is defined as him correctly guessing Alice's bit value. 
The simplest way for Bob to cheat in any Rabin oblivious transfer protocol, whether classical or quantum, is to follow the protocol honestly and then randomly guess the bit value whenever his outcome is ``no bit".  We will call this the guessing strategy. This type of cheating can never be prevented. In an ideal protocol, Bob is not able to cheat any more often than this.

Since Bob learns Alice's bit value with probability $1-p_{?}$, and Bob's guess, if he does not receive the bit value, is correct with probability $1/2$ (assuming Alice is equally likely to choose both bit values), Bob's guessing probability $B^g_{OT}$ is
\begin{equation} \label{eq:Guessing Bob}
B^g_{OT} =1-\frac{p_?}{2} .
\end{equation}

Bob can achieve a higher success probability if he does not follow the protocol honestly. His optimal cheating strategy in the quantum protocol is to make a minimum-error measurement to distinguish between $\ket{\psi_0}$ and  $\ket{\psi_1}$. The minimum-error measurement for distinguishing between two states $\rho_0$ and $\rho_1$, occurring with probabilities $p_0$ and $p_1$, is a projective measurement in the eigenbasis of  $p_0\rho_0-p_1\rho_1$~\cite{Helstrom76}. Bob's cheating probability $B^q_{OT}$ is
\begin{equation} \label{eq:Cheating Bob}
B^q_{OT} = \frac{1}{2}\left(1+\sqrt{1-p_?^2}\right) .
\end{equation}
When $p_?=1$, then $\theta = 0 \degree$ and $\ket{\psi_0} = \ket{\psi_1}$, and $B^q_{OT} = 1/2$; he cannot do any better than with a random guess. When $\theta = 45 \degree$ and the two states are orthogonal, $B^q_{OT} = 1$, but in this case $p_?=0$ and he would also always receive the bit value if he is honest. If $p_{?}=1/2$, corresponding to $\theta = 30 \degree$, Bob's cheating probability is relatively high at $B^q_{OT} = \big(2+\sqrt{3} \big)/4 \approx 0.933$; his guessing probability is $3/4$.
Bob's cheating and guessing probabilities coincide when $p_? = 0$ or $p_? = 1$. In between, Bob's cheating probability is always higher than his guessing probability, as shown in Fig. \ref{fig:A_OT plots}.

\section{Dishonest Alice} \label{alice cheats}

Cheating by Alice is usually defined as her wanting to know if Bob has received the bit or not.
This is the most common definition for Alice cheating in a Rabin OT protocol. A less common definition is for a dishonest Alice to not only want to learn if Bob has received a bit or not, but also the value of the received bit. In classical Rabin OT protocols, a situation where Alice does not know the value of the bit Bob received does not arise. When she sends a bit, she knows its value, no matter if she is cheating or not. In quantum protocols, however, when Alice is dishonest and deviates from the protocol, it can happen that she does not know what bit value Bob thinks he has received. A dishonest Alice deviating from the protocol can send any state, and this state does not need to correspond to a specific bit value. Bob's measurement results are probabilistic, and Alice generally cannot predict what result he will obtain. Thus, Alice will always know what quantum state she sends, but not necessarily Bob's measurement outcome, meaning that she may be uncertain about the bit value Bob receives when he does receive it.
We focus mostly on the more common definition of a dishonest Alice, but we also include a cheating strategy for a dishonest Alice who also wants to learn what bit value Bob thinks he has received.

Alice can always cheat by following the protocol honestly and randomly guessing whether or not Bob received the bit value. In an ideal protocol, she is not able to cheat any more often than this. Alice knows that Bob receives no bit with probability $p_{?}$, and receives the bit with probability $1-p_{?}$.
She will guess the most likely outcome. Hence, Alice's guessing probability $A^g_{OT}$ is 
\begin{align} \label{eq:Guessing Alice}
A^g_{OT} = \max(p_?, 1-p_?) .
\end{align}
Alice will also know which value Bob's received bit would have, since she followed the protocol.
When $p_? = 1$ or $p_? = 0$, Bob will never, respectively always, receive the bit, and necessarily $A^g_{OT}=1$. When $p_?=1/2$, corresponding to $\theta = 30 \degree$, Alice's guessing probability takes its smallest value $A^g_{OT} = 1/2$.

The guessing strategy is usually not Alice's optimal cheating strategy. Her optimal strategy in the quantum protocol depends on whether and how Bob tests the states Alice sends to check if she cheats. In the following subsections, we outline Alice's cheating strategies and probabilities $A^q_{OT}$, first when Bob does no testing at all, and then for two different testing methods. We plot the cheating probabilities in Fig. \ref{fig:A_OT plots}.

\subsection{No testing by Bob}

If Bob does not do any testing in the quantum protocol, Alice can send whatever state suits her best. Her best choice is to send the state $\ket{1}$. Since this is orthogonal to Bob's inconclusive measurement operator (Eq. \eqref{Eq: Bob Measurement Operators}), he will always receive a bit value. Thus, Alice can cheat perfectly, $A^q_{OT} = 1$. Note that she then has no information about what bit value Bob would receive.

\subsection{Bob testing -- Monitoring the probabilities}

One testing strategy that Bob can apply, if the protocol is repeated many times, is to check that the probabilities of obtaining the bit values 0 or 1, or of obtaining no bit, are what he expects them to be. Bob does not discard any states, and does not ask Alice for any more information.
As we will now show, Alice's best cheating strategy is then to send a statistical mixture of the states $\ket{0}$ and $\ket{1}$. Bob's measurement operators for the cases ``bit" and ``no bit" are
\begin{eqnarray}
\Pi_\text{bit} = \Pi_0 + \Pi_1 &=& \text{tan}^2\theta \ket{0}\bra{0}+\ket{1}\bra{1},\nonumber\\
\Pi_? &=& (1-\text{tan}^2\theta) \ket{0}\bra{0}.
\end{eqnarray}

In order for Alice to maximise Bob's probability of obtaining a particular outcome (in this case, ``bit" or ``no bit"), it is optimal for her to send Bob the pure state, which is an eigenstate of Bob's corresponding measurement operator, with the highest eigenvalue. She therefore will not benefit from entanglement.
As we already saw, Alice can make the probability of Bob receiving a bit $p_{\rm bit}=1$ by sending him the state $\ket 1$. Bob will then receive the two bit values with probability 1/2 each. Bob, however, expects to not receive a bit value with probability $p_?$. Alice can make $p_?$ take its maximum value $1-\tan^2\theta$ by sending Bob the state $\ket 0$; Bob will then receive the two bit values with equal probability when he nevertheless does receive a bit value.

If $1-\tan^2\theta\ge 1/2$, meaning that $0\le \theta \le \arcsin \left(1/\sqrt{3} \right) \approx 35.264 \degree$, i.e., $p_?\geq1/3$, Alice guesses that Bob receives no bit if she sent him $\ket 0$ (and that he receives the bit if she sent him $|1\rangle$); if $1-\tan^2\theta<1/2$, her guess is always that Bob receives the bit. 
When $1-\tan^2\theta=1/2$, she can use either guessing pattern.

Bob expects $p_?=\cos(2\theta)$. Thus, if $x$ is the probability for a cheating Alice to send $|0\rangle$, it must hold that $x (1-\tan^2\theta) = \cos(2\theta)$.
It follows that Alice's optimal cheating strategy is to send $\ket 0$ with probability $x= \cos^2\theta$ and $\ket 1$ with probability $1-x= \sin^2\theta$. 
Her probability to correctly guess whether or not Bob receives a bit is then
\begin{align}\label{eq:A_OT Bob monitoring}
A^q_{OT} &= \max[(1+p_?)/2, 1-p_?] .
\end{align}
When $1-p_?>(1+p_?)/2$, that is, when $p_?<1/3$, Alice might also follow the protocol honestly and still achieve the same overall cheating probability $A^q_{OT}=1-p_?$. That is, her cheating probability is the same as in an ideal protocol. By sometimes sending $\ket 1$, however, Alice can be sure that when she does so, Bob will think that he received a bit (as before, at the expense of Alice not knowing what the bit value is). In an ideal protocol, Alice would never know for sure whether Bob has received the bit or not. Arguably, it matters not only what Alice's probability to guess correctly is, but also whether she sometimes can know for sure that her guess is correct.

In a Rabin OT protocol, Bob will usually receive the bit with probability $1/2$, corresponding to $\theta=30\degree$. In this case, $A^q_{OT} = 3/4$, and Alice will need to send $\ket{0}$ $3/4$ of the time, guessing that Bob did not receive the bit, and $\ket{1}$ the rest of the time, guessing that Bob received the bit.

\subsection{Full testing by Bob}

We now consider that Bob tests the states in a way similar to what the receiver does in the 1-out-of-2 oblivious transfer protocol by Amiri \textit{et al.}~\cite{Amiri21}. It is then necessary to implement not just one round of the protocol, but $N$ rounds, and some of the rounds will be used for tests and discarded.
Alice transmits $N$ states to Bob who randomly picks a small fraction $F$ of them to test, where $0 < F \ll 1$. He asks Alice to reveal what these states are, and tests her declarations by measuring the qubits in the appropriate bases. If any of his measurement outcomes do not agree with Alice's declaration, the protocol is aborted. Otherwise, Bob discards the states used for testing and proceeds with the regular protocol using the remaining $N(1-F)$ states. We will refer to this testing method as the full testing scheme. In principle, this could lower Alice's cheating probability as compared with Bob monitoring only the probabilities of his outcomes as in the previous section. We will show that somewhat counterintuitively, Alice's cheating probability remains the same. 

Alice does, however, now need to prepare entangled states in order to cheat.
To always pass Bob's test, Alice needs to send a superposition of the two states she would send when honest, entangled with a system she keeps on her side. This is a state of the form
\begin{equation} \label{eq:Alice general cheat state, full test}
\ket{\Psi_{\text{cheat}}} = a \ket{0}_A \ket{\psi_0} + b \ket{1}_A \ket{\psi_1} ,
\end{equation}
where $\{\ket{0}_A, \ket{1}_A\}$ is an orthonormal basis for the system Alice keeps, $a, b \in \mathbb{R}$, and $|a|^2 + |b|^2 = 1$. The parameters $a$ and $b$ can be chosen real and positive with no loss of generality, since any phase factors can always be absorbed into the kets $\ket{0}_A, \ket{1}_A$. The state $\ket{\Psi_{\text{cheat}}}$ allows Alice to always declare a correct state for Bob's test, by measuring her system in the $\{\ket{0}_A, \ket{1}_A\}$ basis.

Bob still expects to receive the bit values $0$ and $1$ with equal probability.
This would seem to imply that we need $|a|=|b|$ in Eq. \eqref{eq:Alice general cheat state, full test}. Alice could, however, alternate between sending the state in Eq. \eqref{eq:Alice general cheat state, full test}, and an analogous state where $a$ and $b$ are interchanged, so that Bob's probabilities for the two bit values are equal on average. Because of symmetry, her cheating probability would be the same in either case. Therefore, we can still assume a state of the form given in Eq. \eqref{eq:Alice general cheat state, full test}.

\subsubsection{Alice guessing only whether Bob received a bit or not}

After Bob's measurement on his part of $\ket{\Psi_{\text{cheat}}}$, Alice's system will be in one of two possible states, depending on whether Bob received a bit or not. She will need to distinguish between
\begin{equation}
\rho_{\text{no bit}} = 
\begin{pmatrix}
|a|^2 & ab^* \\
a^*b & |b|^2
\end{pmatrix} \text{ \ and \ \ }
\rho_{\text{bit}} = 
\begin{pmatrix}
|a|^2 & 0 \\
0 & |b|^2
\end{pmatrix} ,
\end{equation}
corresponding to Bob not receiving and receiving the bit, respectively. The first case occurs with probability $\cos(2\theta)$ and the second with probability $2\sin^2\theta$. Alice's optimal cheating strategy is to distinguish between these two states with a minimum-error measurement~\cite{Helstrom76}.
Her cheating probability $A^q_{OT}$ is then
\begin{align}
A^q_{OT} &= \max[(1+ u)/2 , 1-p_?] , 
\end{align}
where $u=\sqrt{(1-2p_?)^2 -4a^2b^2 (1-p_?) (1-3p_?)}$.
Since $u \le 1-2p_?$, it is clear that when $p_?\le 1/3$, Alice cannot cheat any better than with a random guess. When $p_? \geq 1/3$, Alice's optimal choice for the coefficients $a$ and $b$ is $a=b=1/\sqrt{2}$, which maximises $u$, meaning that she needs to distinguish between
\begin{align}
\rho_{\text{no bit}} &= 
\begin{pmatrix}
1/2 & 1/2 \\
1/2 & 1/2
\end{pmatrix} = \ket{+} \bra{+} , \nonumber \\
\rho_{\text{bit}} &= 
\begin{pmatrix}
1/2 & 0 \\
0 & 1/2
\end{pmatrix} = \dfrac{1}{2} \mathbb{1} .
\end{align}
Her cheating probability becomes
\begin{align} \label{eq: Alice cheats optimum standard, Bob full testing}
A^q_{OT} &=  \max[(1+p_?)/2, 1-p_?] .
\end{align}
As before, Alice can cheat perfectly when $p_? = 0$ or $p_? = 1$. In these cases, Bob always receives a bit, or never receives the bit, respectively.
For the usual choice of $p_?=1/2$, we have $A^q_{OT} = 3/4$. 
Alice's cheating probability reaches its minimum, $A^q_{OT} = 2/3$, when $p_?= 1/3$.
When $p_? \leq 1/3$, Alice's cheating probability coincides with her guessing probability, which is what should hold in an ideal protocol. Bob can, however, still cheat with a probability larger than his guessing probability.

Alice's optimal cheating probability when Bob does the full testing, Eq. \eqref{eq: Alice cheats optimum standard, Bob full testing}, is the same as when Bob monitors the probabilities, Eq. \eqref{eq:A_OT Bob monitoring}.
This might at first seem surprising, but can be explained as follows. When it is not a testing round, Alice cheats by measuring her system in the $\{\ket{+}_A, \ket{-}_A\}$ basis. When $a=b=1/\sqrt{2}$, the states on Bob's side then become $\ket{0}$ or $\ket{1}$, the same states Alice sends in the situation where Bob is monitoring the occurrence probabilities. Thus, full testing does not help Bob against an Alice who can prepare entangled states, and he can just as well monitor only the probabilities, with the added benefit that no states need to be discarded. It is also intuitively clear that Bob can benefit from the full tests if Alice cannot prepare entangled states, and/or cannot store her half of the entangled state until she knows whether Bob is testing her or not. Using the full testing strategy, Bob can then prevent Alice from cheating any better than with a random guess, since she then has to follow the protocol honestly in order to pass Bob's full tests. This is consistent with the fact that if there are restrictions on quantum memory \cite{Damgard08}, then perfect quantum oblivious transfer is possible (also Bob can be prevented from cheating if his quantum memory is restricted).

\subsubsection{Alice guessing also Bob's bit values}

If Alice wants to in addition also know what Bob's bit value is when she is cheating, then this will change her cheating strategy and cheating probability.
After Bob's measurement on $\ket{\Psi_{\text{cheat}}}$, Alice's system will be in one of three possible states, depending on Bob's measurement outcome. She will need to distinguish between
\begin{equation} \label{General 3 states}
\rho_{\text{no bit}} =
\begin{pmatrix}
|a|^2 & ab^* \\
a^*b & |b|^2
\end{pmatrix}, 
\rho_{\text{bit }0} = 
\begin{pmatrix}
1 & 0 \\
0 & 0
\end{pmatrix}, 
\rho_{\text{bit }1} = 
\begin{pmatrix}
0 & 0 \\
0 & 1
\end{pmatrix},
\end{equation}
corresponding to Bob not obtaining the bit, or Bob obtaining bit value $0$ or bit value $1$, respectively. The corresponding probabilities are $\cos(2\theta)$ for $\rho_{\text{no bit}}$, $2|a|^2\sin^2\theta$ for $\rho_{\text{bit }0}$, and $2|b|^2\sin^2\theta$ for $\rho_{\text{bit }1}$.

It is impossible for Alice to have unambiguous knowledge of both that Bob received the bit, and its value. Alice can only have unambiguous knowledge of Bob's possible bit value if she measures her system in the $\{\ket{0}_A, \ket{1}_A\}$ basis, corresponding to following the protocol honestly, and then she is left with a random guess of whether Bob receives the bit or not. Alice can, however, make a minimum-error measurement to distinguish between the three states in Eq. \eqref{General 3 states}.
We can bound Alice's cheating probability in this case by assuming that she performs a square-root measurement (SRM)~\cite{Hausladen94}. This is not necessarily her optimum strategy. In fact, it turns out that for some values of $\theta$ and $a$, she even does better using her guessing strategy. But the SRM often performs reasonably well, and will provide a lower bound on Alice's cheating probability. The lower bound given by the SRM can be shown to be 
\begin{eqnarray} \label{eq: Alice cheats 3 states Bob tests}
A^q_{OT}&\geq& \Big\{ \left(1-p_?+2p_?^2 \right)\left(1+2ab\sqrt{\frac{1-p_?}{1+p_?}}\right)\\
&&+p_?\left( 6p_?a^2b^2\frac{1-p_?}{1+p_?}-1\right)\Big\} \Big/\left[1+2ab\sqrt{1-p_?^2}\right].\nonumber
\end{eqnarray}

Substituting $b^2=1-a^2$ and looking at the partial derivative of Eq. \eqref{eq: Alice cheats 3 states Bob tests} with respect to $a$, we can show that Alice's optimal choice for the constants $a$ and $b$, when she makes the SRM, is $a = b = 1/\sqrt{2}$. Substituting these values into Eq. \eqref{General 3 states}, the three states Alice wants to distinguish between, become
\begin{align} \label{Mirror Symmetric States}
\rho_{\text{no bit}} &=
\begin{pmatrix}
1/2 & 1/2 \\
1/2 & 1/2
\end{pmatrix}  = \ket{+} \bra{+} , \nonumber \\
\rho_{\text{bit }0} &= 
\begin{pmatrix}
1 & 0 \\
0 & 0
\end{pmatrix} = \ket{0} \bra{0}  , \nonumber \\
\rho_{\text{bit }1} &= 
\begin{pmatrix}
0 & 0 \\
0 & 1
\end{pmatrix} = \ket{1} \bra{1} ,
\end{align}
occurring with probability $\cos(2\theta)$ for $\rho_{\rm no\ bit}$ and probability $\sin^2\theta$ for each of $\rho_{\text{bit }0}$ and $\rho_{\text{bit }1}$.
These states are mirror-symmetric; the unitary operation for which $U\ket{0}=\ket{1}$ and $U\ket{1}=\ket{0}$ keeps $\rho_{\text{no bit}}$ unchanged and interchanges $\rho_{\text{bit }0}$ and $\rho_{\text{bit }1}$. The optimal minimum-error measurement is known for mirror-symmetric states~\cite{Andersson02}, and we no longer need to use the square-root measurement when $a=b$.
Alice's cheating probability is then
\begin{align}
A^q_{OT} &= \max [4p^2_?/(5p_?-1) , 1-p_?] .
\end{align}
We conjecture that choosing $a=b$ is in fact optimal for Alice.
As in the standard cheating scenario, where dishonest Alice's aim is to distinguish between Bob not receiving the bit and him receiving the bit, Alice can perfectly know Bob's outcome when $p_?=0$ or $p_?=1$, but in these cases Bob also either never receives the bit or always receives it. When $p_?=2/5$, Alice's cheating probability is at its minimum with $A^q_{OT} = 16/25 = 0.64$. For $p_?=1/3$ and $p_?=1/2$, we have $A^q_{OT} = 2/3$. Furthermore,  whenever $p_? \leq 1/3$, Alice's cheating probability coincides with her guessing probability, as was also the case for all of the other cheating scenarios above.

\begin{figure}[h]
\centering
\includegraphics[scale=0.5]{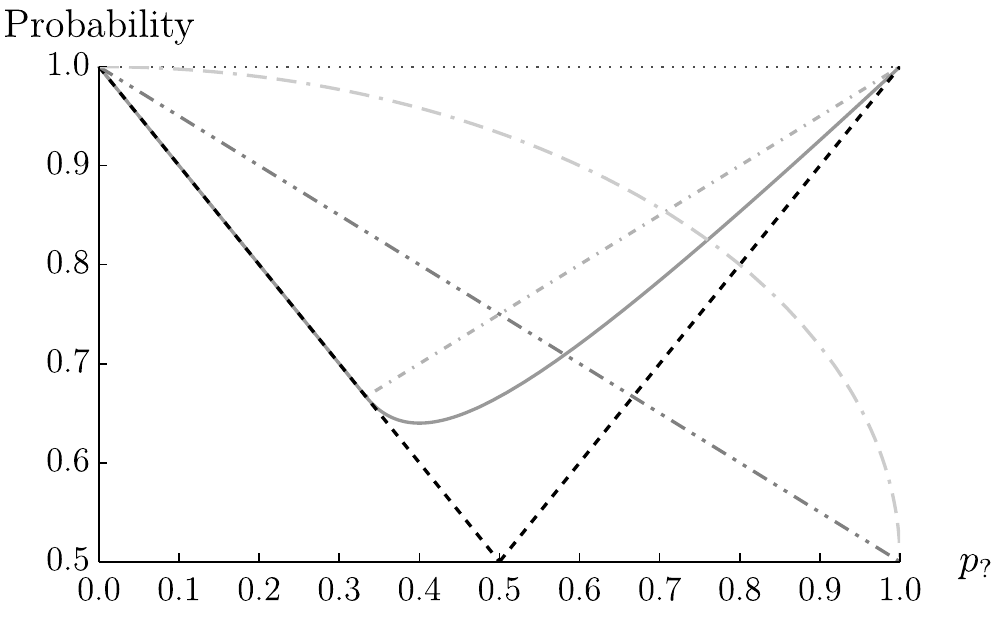}
\caption{Alice's and Bob's cheating probabilities for the different cheating scenarios as functions of $p_?$. Alice's guessing probability is plotted as the black dashed line.  The dotted line shows Alice's cheating probability when Bob does not do any testing. Her cheating probability when Bob does the full testing scheme (for the optimal case with $a = b = 1/\sqrt{2}$), or when Bob monitors the probabilities, is shown by the dash-dotted line. Alice's cheating probability when Bob does the full testing and Alice wants to distinguish between outcomes ``no bit", ``bit of value $0$", and ``bit of value $1$", is given by the solid line (for the conjectured optimal case with $a = b = 1/\sqrt{2}$). Bob's guessing probability is shown by the dot-dot-dashed line, and his cheating probability by the dash-dash-dotted line.}
\label{fig:A_OT plots}
\end{figure}

Alice's guessing probability and her different cheating probabilities discussed in this section, as well as Bob's guessing and cheating probabilities, are plotted in Fig. \ref{fig:A_OT plots} as functions of $p_?$.
For any kind of testing by Bob, it holds that when $p_? \leq 1/3$, Alice's cheating probability is equal to the guessing probability. However, using the cheating strategies might give Alice extra knowledge; for instance, if she sends $\ket{1}$ when Bob is monitoring the occurrence probabilities, she is certain that Bob will receive a bit value.
The figure also shows that, when Bob does the full testing, it is harder for Alice to cheat if she is also trying to obtain information about the received  bit values, than if she only tries to obtain information about whether Bob received a bit value or not (except then when a random guess is Alice's best cheating strategy).

\section{Evaluation of the protocol} \label{sec: Comparison}

To evaluate the performance of the quantum Rabin OT protocol using two pure states, we compare it to an ideal Rabin OT protocol where both parties can cheat no better than with a random guess, and to classical Rabin OT protocols. We also compare the quantum Rabin OT protocol to the protocols by Bansal and Sikora~\cite{Bansal23}, for which $p_?=1/2$.

\subsection{Comparison to classical Rabin OT protocols} \label{sec: Comparison Classical}

The considered pure-state quantum Rabin OT protocol makes use of only two parties, a sender and a receiver. Thus, we focus on classical Rabin OT protocols of a comparable set-up, i.e., protocols that are also not based on the use of a third party.
We also assume here that there are no guaranteed errors, since a noisy channel with guaranteed errors is a resource that can be used to construct secure oblivious transfer~\cite{Damgard04}.
There are two ways to construct such classical Rabin  oblivious transfer protocols.
One way is to create a protocol where Alice sends a bit with some probability $s$ and Bob reads the bit value with some probability $r$, leading to an overall probability of $rs$ for Bob to receive the bit value. The second way is to build a protocol by probabilistically choosing between two trivial classical Rabin OT protocols, where either Alice or Bob can cheat perfectly, while the other party can only cheat with a random guess.

We show in Appendix \ref{App: Classical coin flip protocol} that the classical Rabin OT protocols using the probabilities $s$ and $r$ perform better than the classical Rabin OT protocols based on a probabilistic choice between trivial protocols. Hence, it is sufficient to compare the quantum protocol to the former type of classical protocols only; any quantum advantage can only be achieved when none of the classical protocols is better.
Note also that this first type of protocols can be generalised further. A protocol can be constructed by a probabilistic choice between $k$ different protocols, where each of them occurs with a probability $p_k$ respectively, and has probabilities $s_k$ and $r_k$, with meanings analogous to above. As we show in Appendix \ref{App: Generalisation of random factor protocol}, the performances of the first type of protocol and its generalisation coincide. Thus, we can, without loss of generality, focus on the protocol with one set of probabilities $s$ and $r$ in the subsequent analysis.
A classical protocol of this type with one set of probabilities $s$ and $r$ is defined as follows.

\begin{definition} \label{Def: Classical rs protocol}
Alice holds a bit, with the values $0$ or $1$ equally likely. She sends the bit to Bob with probability $s$. With probability $r$, Bob attempts to read the bit value.
\end{definition}

\noindent In this protocol, the probability of an honest Bob not receiving the bit is $p_? = 1-rs$.
When Alice is dishonest, she can cheat perfectly by never sending Bob the bit and she will then know that he will not receive the bit. Hence, we need to include some testing by Bob. That is, he will need to check that, on average, he receives the bit with the expected probability $p_\text{bit}=rs$. This will force Alice to send the bit with probability $s$, bounding her average cheating probability. If cheating, Alice will guess whatever is most likely, taking into account whether or not she sent Bob the bit, and with what probability an honest Bob will attempt to read the bit.
A dishonest Bob will always attempt to read the bit, and if he did not receive a bit from Alice, then he will guess its value.
Alice's and Bob's cheating probabilities for this classical Rabin OT protocol are therefore
\begin{align} \label{eq:Alice classical cheating}
A^{c}_{OT} &=
\begin{cases}
1-rs = p_? &\text{ for } r < 1/2 \\
1-s+rs = 2 - p_? -s &\text{ for } r \geq 1/2, \end{cases} \\
B^{c}_{OT} &= \frac{1}{2} (1+s) . \label{eq:Bob classical cheating}
\end{align}

For $p_? \leq 1/2$, we need $r \geq 1/2$ and Alice's cheating probability is described by the second case in Eq. \eqref{eq:Alice classical cheating}. When $p_? > 1/2$, then both cases, $r < 1/2$ and $r \geq 1/2$, can hold. The crossover between the two cases in Eq. \eqref{eq:Alice classical cheating} happens when $p_? = 1 - s/2$. Thus, Alice's cheating probability is described by $A^{c}_{OT} = p_?$ for $1/2 \leq 1 - s/2 < p_?$ and by $A^{c}_{OT} = 2 - p_? -s$ for $1/2 < p_? \leq 1 - s/2$. Note that whether Alice's cheating probability for a given value of $p_? > 1/2$ is specified by the first or by the second case in Eq. \eqref{eq:Alice classical cheating} is entirely dependent on the values of the variables $r$ and $s$.
Substituting $s = 2B^{c}_{OT} -1$ into the later expression for $A^{c}_{OT} $ and rearranging, we obtain a tradeoff relation
\begin{equation} \label{eq: Tradeoff random factor}
B^{c}_{OT} = \dfrac{1}{2} \big(3 - p_? -  A^{c}_{OT} \big)
\end{equation}
between Alice's and Bob's cheating probabilities.
In this case, that is when $r \geq 1/2$, fixed values for $p_?$ and for $A^{c}_{OT}$ determine the value of $B^{c}_{OT}$. 
For $r < 1/2$, fixing the value of $p_?$ also fixes Alice's cheating probability $A^{c}_{OT}$, while $B^{c}_{OT}$ can be variable to some extent, depending on the values for $r$ and $s$. In fact, there are values for $r$ and $s$ that can be picked to achieve the same values of $p_?$ and $A^{c}_{OT}$ as when $r = 1/2$, but with a larger $B^{c}_{OT}$ than given by Eq. \eqref{eq: Tradeoff random factor}.
Therefore, and since we only care about the ``best" protocols with regards to simultaneously lowering Alice's and Bob's cheating probabilities as much as possible for given values of $p_?$, we can say that the protocols with $r < 1/2$ are inferior compared to the ones with $r \geq 1/2$. An honest Alice and an honest Bob would never agree on a protocol with $r < 1/2$ and we can, without loss of generality, solely focus on the region $r \geq 1/2$ in our comparison of the different Rabin OT protocols.

\begin{figure}[h]
\centering
\includegraphics[scale=0.4]{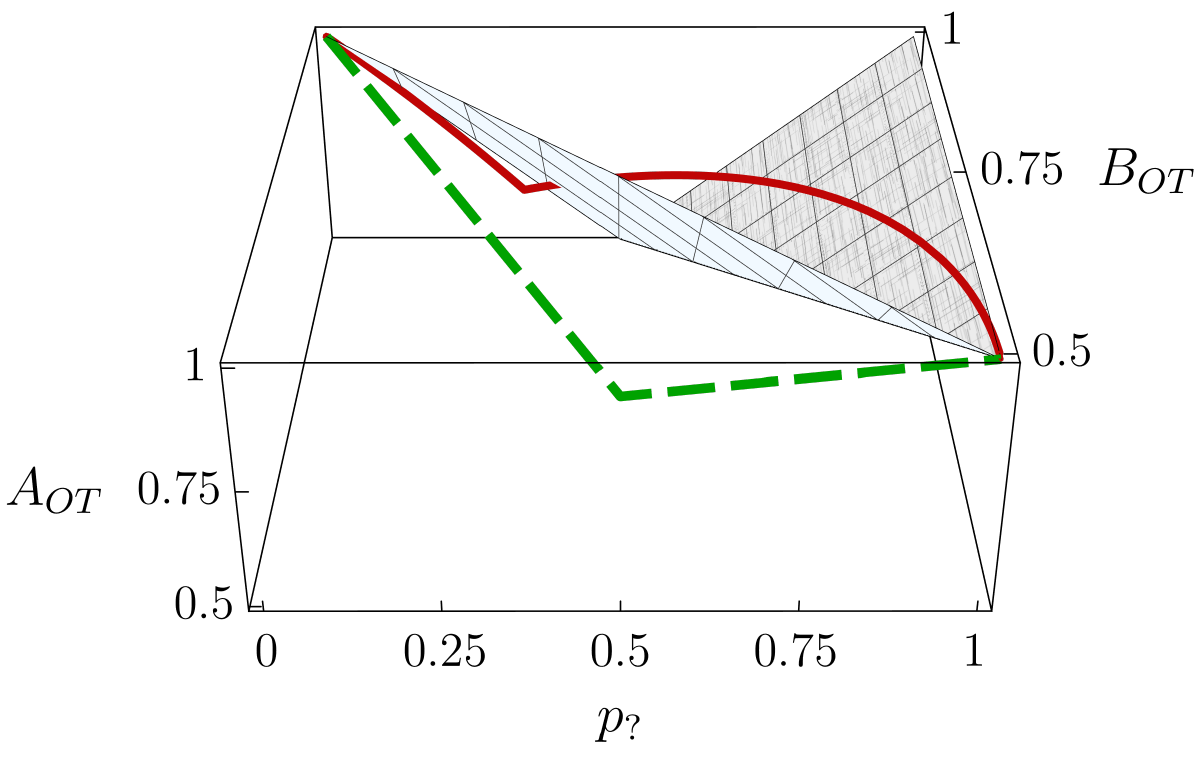}
\caption{Plot showing Alice's and Bob's cheating probabilities for different values of $p_?$. The dashed green line illustrates the ideal protocol, the blue plane the classical protocol when $r \geq 1/2$, the shaded gray plane the classical protocol when $r < 1/2$, and the red line the quantum protocol.}
\label{fig:Comparisons plot}
\end{figure}

Using $s = 2B^{c}_{OT} -1$ and $r = (1-p_?)/(2B^{c}_{OT} -1)$, we can plot the cheating probabilities in relation to each other for each $p_?$. In the same plot, we include the cheating probabilities of an ideal protocol (Eqns. \eqref{eq:Guessing Bob} and \eqref{eq:Guessing Alice}) and of the quantum protocol using pure states (Eqns. \eqref{eq:Cheating Bob} and \eqref{eq:A_OT Bob monitoring}).
The green dashed line in Fig. \ref{fig:Comparisons plot} shows the cheating probabilities for Alice and Bob for each $p_?$ in an ideal Rabin OT protocol. For the classical protocol, we can see that $A^{c}_{OT}$ and $B^{c}_{OT}$ can take on various combinations of values for most values of $p_?$. This range arises from the different values of $r$ and $s$, which can lead to the same value of $p_?$, but with different cheating probabilities for Alice and Bob.  The classical protocol is illustrated by the blue plane when $r \geq 1/2$, and by the shaded gray plane when $r < 1/2$.
We note that the shaded gray plane is just a projection back in the direction of the $B_{OT}$-axis along the edge of the blue plane. This confirms our earlier observation that protocols with $r < 1/2$ attain larger values for $B^{c}_{OT}$ than protocols with the same values of $p_?$ and $A^{c}_{OT}$ that sit on the edge of the blue plane where $r = 1/2$.
The quantum protocol is visualised by the red line. It only has a quantum advantage for $0<p_? < 5/13$, which is where the line for the quantum protocol is slightly below and in front of the planes for the classical protocol, and thus closer to the line for the ideal protocol.

Note that the classical protocol can be expressed as a quantum protocol where from Bob's point of view, each of Alice's bit values corresponds to a mixed state sent by an honest Alice, and an honest Bob makes a generalized quantum measurement (he either makes a projective measurement, or does nothing); see Appendix \ref{App: Mixed States Protocol} for more details.
Considering that this classical (or mixed-state quantum) protocol is better than the quantum protocol using pure states for a range of values of $p_?$, we conclude that mixed quantum states can sometimes give lower cheating probabilities in quantum Rabin oblivious transfer protocols, which merits further investigation. It is possible that cheating probabilities could be further lowered in more general mixed-state quantum protocols.

\subsection{Comparison to another quantum Rabin OT protocol}

As mentioned earlier, two quantum Rabin OT protocols based on stochastic selection have been proposed by Bansal and Sikora~\cite{Bansal23}. These protocols are defined for $p_?=1/2$ only.
In both these stochastic switching protocols, the sender Alice sends a qutrit state to Bob who then performs a coin flip in order to select which one of two possible ``bad" Rabin OT protocols is executed. One of the ``bad" protocols is a straightforward execution of a Rabin OT protocol and the other one includes a testing step by Bob where Alice needs to declare the bit's value, Bob tests her declaration, and then they restart the protocol without testing when Alice passes the test.
The security analysis of one of the stochastic switching protocols is based on a different and less common definition for a dishonest Alice. That is, a dishonest Alice is defined as her wanting to force Bob's output to be ``no bit".
The security analysis of the other stochastic switching protocol is based on the same definition for a dishonest Alice as we use in this paper. That is, a dishonest Alice wants to correctly guess whether Bob has received the bit or not. Hence, we can compare this stochastic switching protocol directly to the quantum Rabin OT protocol we are analysing.

It is interesting to note that both ``bad" protocols in~\cite{Bansal23} have the same cheating probabilities as the pure-state protocol which we have analysed, has for $p_?=1/2$, when no testing by Bob is applied; $A^q_{OT}=1$ and $B^q_{OT} \approx 0.933$. In~\cite{Bansal23}, the overall security is improved by the stochastic selection, and the switching protocol achieves cheating probabilities $A^{\text{BS}}_{OT} = 0.933$ and $B^{\text{BS}}_{OT} = 0.9691$. Thus, we can see that Alice's cheating probability is lowered, while Bob's cheating probability increases. In comparison, by adding some testing by Bob, we can improve the security of the pure-state protocol we have analysed, lowering Alice's cheating probability while keeping Bob's cheating probability the same. Therefore, we can conclude that the pure-state protocol analysed here outperforms the switching protocol in~\cite{Bansal23}.

\begin{figure}[h]
\centering
\includegraphics[scale=0.35]{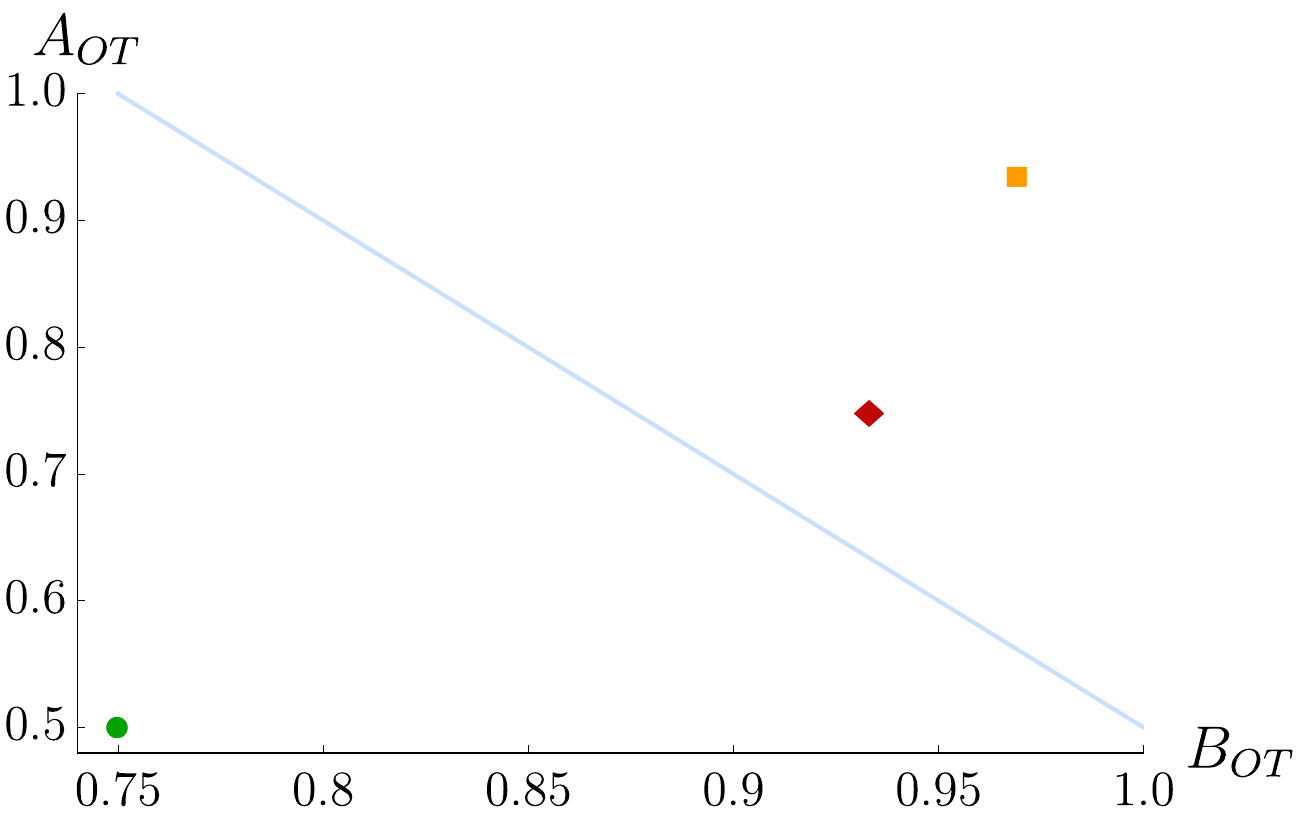}
\caption{Plot showing Alice's and Bob's cheating probabilities for $p_? = 1/2$ in different Rabin oblivious transfer protocols. The green circle corresponds to the ideal protocol, the blue line to our classical protocol, the red diamond to the quantum protocol based on two pure states, and the orange square to the stochastic switching protocol by Bansal and Sikora~\cite{Bansal23} with cheating probabilities $A^{\text{BS}}_{OT} = 0.933$ and $B^{\text{BS}}_{OT} = 0.9691$.}
\label{fig:Comparisons plot with other protocols}
\end{figure}

This is illustrated in Fig. \ref{fig:Comparisons plot with other protocols}, where we plot cheating probabilities for both quantum protocols, the classical protocol, and the ideal protocol, in all cases for $p_?=1/2$. The plot shows that the switching protocol by Bansal and Sikora (orange square) has higher cheating probabilities for both Alice and Bob than the quantum protocol  we have analysed (red diamond) and is thus obviously also further away from the classical protocol we proposed and analysed (blue line) and the ideal protocol (green circle).

\section{Conclusions}

We have examined a quantum Rabin oblivious transfer protocol proposed by Cheong \textit{et al.}~\cite{Cheong10}, where a sender Alice encodes her bit values using two pure states with overlap $\cos(2\theta)$, and a receiver Bob performs an unambiguous state discrimination measurement. This perfectly distinguishes between Alice's states with a probability $1-p_?=1-\cos(2\theta)$, which therefore is the probability that Bob receives Alice's bit value.

Usually one is interested in the sender's and receiver's cheating probabilities. Cheong \textit{et al.}~\cite{Cheong10} did consider Bob's cheating probability, but instead of Alice's cheating probability, they looked at her ``advantage", which they defined as how much Alice can decrease or increase Bob's probability to receive a bit value. 
We have instead examined Alice's cheating probability, which is a more commonly used figure of merit. If Bob does not test the states Alice sends, then she can in fact cheat perfectly. One can then argue that the quantum protocol is worse than a trivial classical protocol where Alice sends Bob a bit with probability $1-p_?$, and does not retain a record of what she has done if she is honest. In both protocols, Alice can cheat perfectly. In the trivial classical protocol, Bob cannot cheat any better than with a random guess, but in the quantum protocol he can cheat more often than this.

However, as we have shown, Alice's cheating probability can be lowered if Bob tests the states Alice sends. In return, Alice's cheating probability becomes an average cheating probability.
We also looked at how well Alice can cheat if she not only wants to distinguish between Bob receiving and not receiving the bit, but also wants to know which of the two bit values he has obtained if he does receive a bit. This is harder for Alice to do, and her cheating probability can be lowered further.
Note that this different definition of what a cheating Alice aims to learn is not applicable in the classical setting. When Alice (dishonest or not) sends a bit in a classical protocol, she always knows its value, whereas in a quantum protocol, a dishonest Alice always knows what quantum state she sends, but not necessarily what bit value Bob thinks he receives, because of the probabilistic nature of Bob's measurement outcomes.
Even if the definition of a dishonest Alice wanting to also learn the value of the bit Bob has received is not much considered in the literature, we deem it an important aspect for quantum Rabin OT protocols. While the value of the bit is given knowledge for an honest as well as a dishonest Alice in classical protocols, it is additional  information for a dishonest Alice in quantum protocols.  It is more difficult for a dishonest Alice  to learn both whether Bob has received the bit or not, and its value if it was received.
Which of these cheating definitions is more proper, is not completely clear and it may depend on the application. This question merits further investigation.

It turned out that for $p_?\le 1/3$, all of Alice's optimal cheating probabilities (except when Bob does not do any testing and she can cheat perfectly) are equal to her guessing probability. This is the probability of her cheating by following the protocol honestly and making a random guess, which is the lowest cheating probability for Alice that can be achieved even in an ideal protocol. Honestly following the protocol means that Alice will know with certainty which value the received bit would have. If she does not follow the protocol honestly but sends $\ket{1}$ as often as possible, then she will be certain that Bob did receive a bit when she did send $\ket 1$, but will be unsure about Bob's bit value.
Either way, the resulting quantum protocol can in this range be said to be better than a trivial classical protocol where Alice sends Bob a bit, and he reads its value with probability $1-p_?$. In both protocols, Alice can only cheat as well as with a random guess, but in the classical protocol Bob can cheat perfectly, while in the quantum protocol, his cheating probability is strictly less than 1.
We also compared the quantum Rabin OT protocol using two pure states to a quantum Rabin OT protocol based on stochastic switching~\cite{Bansal23}. The protocol analysed here achieves lower cheating probabilities for both Alice and Bob.

Finally, we compared the quantum Rabin OT protocol to classical Rabin OT protocols, which do not rely on the involvement of a third party. The classical protocols are based on honest Alice and honest Bob choosing their actions according to some probabilities $s$ and $r$, respectively.
We showed that there exists a quantum advantage for $0 < p_? < 5/13$, but that the classical  protocol outperforms the quantum protocol for $5/13 < p_? < 1$. This range also includes $p_?=1/2$, the most conventional value for $p_?$ in Rabin oblivious transfer.

Since the considered classical Rabin OT protocol based on probabilistic choices by Alice and Bob can be expressed as a quantum protocol using mixed states, it is clear that using mixed states can sometimes give an advantage, that is, lower cheating probabilities. In other words, it would be interesting to investigate more general quantum Rabin OT protocols that use mixed states instead of pure states, determining the optimal mixed states to use, and whether these protocols achieve a quantum advantage for a greater range of $p_?$.

\section*{Acknowledgements}

This work was supported by the UK Engineering and Physical Sciences Research Council (EPSRC) under Grants No. EP/T001011/1 and EP/R513386/1.

\appendix

\section{Comparison of classical protocols} \label{App: Classical coin flip protocol}

As mentioned, there are two ways to construct a classical Rabin oblivious transfer protocol without relying on a third party.
The protocols, which are based on Alice sending a bit with some probability $s$ and Bob choosing to read the bit value with some probability $r$, are defined by Protocol \ref{Def: Classical rs protocol} in Section \ref{sec: Comparison Classical}. We will now define the other type of protocols, which are constructed by probabilistically choosing between two trivial classical Rabin OT protocols.

The Rabin OT protocols obtained by such a probabilistic choice are ``almost classical", since similar to the procedure in~\cite{Chailloux16}, the outcome of a weak coin flip determines which protocol is implemented. Weak coin flipping means that each party has a favoured outcome that signifies that they ``win", and that each party is unable to increase the probability for their preferred outcome. (In strong coin flipping, the parties are unable to bias the coin in either direction.)
Weak coin flipping is impossible classically with information-theoretic security, but possible using a quantum protocol~\cite{Mochon07}.
The two protocols Alice and Bob choose between are defined as follows.

\begin{definition} \label{Def: Protocol Perfect Cheating Alice}
Alice holds a bit, with the values $0$ or $1$  equally likely. She sends the bit to Bob with probability $1-p_?$. She does not keep a record of what she has done. 
\end{definition}

\begin{definition} \label{Def: Protocol Perfect Cheating Bob}
Alice holds a bit, with the values $0$ or $1$  equally likely. She sends it to Bob, who chooses to read the bit with probability $1-p_?$ and discards it unread the rest of the time.
\end{definition}

Combining Protocol \ref{Def: Protocol Perfect Cheating Alice} and Protocol \ref{Def: Protocol Perfect Cheating Bob} with a weak coin flip to determine which protocol is executed, we obtain the following protocol.

\begin{definition} \label{Def: Protocol Coin Flip}
Protocol \ref{Def: Protocol Perfect Cheating Alice}  is executed with probability $y$ and Protocol \ref{Def: Protocol Perfect Cheating Bob} is executed with probability $1-y$.
\end{definition}

The probability that Bob receives the bit in Protocol \ref{Def: Protocol Coin Flip} is then also equal to $1-p_?$.
In Protocol \ref{Def: Protocol Perfect Cheating Alice}, a dishonest Alice can cheat with probability 1, whereas a dishonest Bob can cheat no better than with the guessing probability $B_{OT}^g=1-p_?/2$.
In Protocol \ref{Def: Protocol Perfect Cheating Bob}, a dishonest Bob can cheat perfectly, and a dishonest Alice can cheat no better than with the  guessing probability $A_{OT}^g = \max(p_?, 1-p_?)$.
Hence, for Protocol \ref{Def: Protocol Coin Flip} (based on a coin flip), Alice's and Bob's cheating probabilities are
\begin{align} \label{eq:Alice coin flip cheating}
A^{\text{cf}}_{OT} &=
\begin{cases}
1-p_?+y p_?
&\text{ for } p_? \le 1/2 \\
p_?  +  y (1- p_?) &\text{ for } p_?>1/2 , \end{cases} \\
B^{\text{cf}}_{OT} &= 1 - \frac{yp_?}{2} . \label{eq:Bob coin flip cheating}
\end{align}

In order to compare Protocol \ref{Def: Classical rs protocol} and Protocol \ref{Def: Protocol Coin Flip} to each other, we will obtain tradeoff relations between $A_{OT}$ and $B_{OT}$ for the different protocols. Broadly speaking, if Alice's cheating probability is higher, then Bob's cheating probability is lower, and vice versa. For a given cheating probability for Alice, we would like Bob's cheating probability to be as low as possible, and vice versa.
Eliminating $y$ from Eqns. \eqref{eq:Alice coin flip cheating} and \eqref{eq:Bob coin flip cheating}, which hold for Protocol \ref{Def: Protocol Coin Flip}, we obtain
\begin{eqnarray}
\label{eq:Tradeoff f1 coin flip}
A_{OT}^{\text{cf}}+2B_{OT}^{\text{cf}}
&=& 3 - p_? \quad\quad\quad\quad 
\text{for } p_? \leq 1/2 , \nonumber
\\
p_?A_{OT}^{\text{cf}}+2(1-p_?)B_{OT}^{\text{cf}}
&=& 1+ (1-p_?)^2 \quad 
\text{for } p_? >1/2, \nonumber \\
\label{eq:Tradeoff f2 coin flip}
\end{eqnarray}
which define tradeoff relations between $A_{OT}^{\rm cf}$ and $B_{OT}^{\rm cf}$ in the ranges $p_?\le 1/2$ and $p_?> 1/2$.

From Eqns. \eqref{eq:Alice classical cheating} and \eqref{eq:Bob classical cheating}, one can obtain similar relations for Protocol \ref{Def: Classical rs protocol}. For the two free parameters, $r$ and $s$, it holds that $1-rs=p_?$. When $r \ge 1/2$, which corresponds to $p_? \le 1-s/2$, we obtain
\begin{equation} 
\label{eq:Tradeoff f1 random factor}
A_{OT}^{c}+2B_{OT}^{c}=3-p_?
\end{equation}
for Protocol \ref{Def: Classical rs protocol}. This is the same expression as the one obtained for Protocol \ref{Def: Protocol Coin Flip} when $p_?\le 1/2$.
Since $p_? \le 1/2$ implies $p_? \le1-s/2$, this means that Protocol \ref{Def: Classical rs protocol} and Protocol \ref{Def: Protocol Coin Flip} are equally good when $p_?\le1/2$.

When $p_? > 1/2$, then either $r<1/2$ or $r \ge 1/2$ could hold. That is, we can have either $p_?>1-s/2$ or $p_?\le 1-s/2$. One way to proceed in order to compare Protocol \ref{Def: Classical rs protocol} and Protocol \ref{Def: Protocol Coin Flip} , when $p_? > 1/2$, is to form
\begin{equation}
\label{eq:P4tradeoff1} 
p_?A_{OT}^{c}+2(1-p_?)B_{OT}^{c}=\begin{cases} 
p_?^2 + (1-p_?) (1+s) \quad \\ \text{for } p_? > 1-s/2\ge 1/2 \\[0.5em]
1 + p_? - p_?^2 - s(2p_?-1) \quad \\ \text{for } 1-s/2\ge p_? >1/2
\end{cases} 
\end{equation} 
for Protocol \ref{Def: Classical rs protocol}. The left-hand side (LHS) in the above equation is the same expression as the LHS in the second tradeoff relation in \eqref{eq:Tradeoff f2 coin flip}, valid for Protocol \ref{Def: Protocol Coin Flip} in the region $p_?>1/2$.
For a fixed value of $p_?$, minimizing $p_?^2+(1-p_?)(1+s)$ will correspond to the smallest possible bound on the cheating probabilities for Alice and Bob, and thus the (in this sense) best protocol of type \ref{Def: Classical rs protocol}, in the range $p_? > 1-s/2$. Recall that $s$ is the probability for Alice sending the bit, and $r$ is the probability of an honest Bob attempting to read Alice's bit, and that $r$ and $s$ are parameters that Alice and Bob agree on; they should evidently agree on parameters that give a protocol that is as ``good" as possible. ``Good" can be defined in many ways, and good for Alice is not the same as good for Bob. Here, our objective is to see whether Protocol \ref{Def: Classical rs protocol} or Protocol \ref{Def: Protocol Coin Flip} can achieve lower cheating probabilities, taking both Alice's and Bob's cheating probabilities into account. To get the ``best" protocol of type \ref{Def: Classical rs protocol}, we should evidently choose $s$ as small as possible, but so that $p_?\ge 1-s/2$ still holds. This gives $s=2(1-p_?)$. Similarly, in order to minimize $1 + p_? - p_?^2 - s(2p_?-1)$, we should choose $s$ as large as possible, but so that $1-s/2\ge p_?$ still holds. This again gives $s=2(1-p_?)$. For this optimal choice of $s$, the relevant expression for Protocol \ref{Def: Classical rs protocol} in the range $p_? > 1/2$ becomes
\begin{equation}
\label{eq:P4tradeoff2}
p_?A_{OT}^{\rm rf}+2(1-p_?)B_{OT}^{\rm rf}=3-5p_?+3p_?^2 \text{ for } p_? > 1/2.
\end{equation}
The expression in the right-hand side (RHS) in Eq. \eqref{eq:P4tradeoff2} can then be compared to the expression $1+(1-p_?)^2$ in the RHS of the second line in Eq. \eqref{eq:Tradeoff f2 coin flip}.
Since $3-5p_?+3p_?^2-1-(1-p_?)^2 = (1-2p_?)(1-p_?) \le 0$, we can conclude that the RHS in Eq. \eqref{eq:P4tradeoff2} is always smaller than the RHS on the second line in Eq. \eqref{eq:Tradeoff f2 coin flip}. Therefore, Protocol \ref{Def: Classical rs protocol} is better than Protocol \ref{Def: Protocol Coin Flip} when $p_?>1/2$, in the sense that the value of $p_?A_{OT}+2(1-p_?)B_{OT}$ is smaller.

To summarise, Protocol \ref{Def: Classical rs protocol}  and Protocol \ref{Def: Protocol Coin Flip} are equally good when $p_?\le 1/2$, and Protocol \ref{Def: Classical rs protocol} is better than Protocol \ref{Def: Protocol Coin Flip} when $p_? > 1/2$, in the sense that if we fix $p_?$ and either Alice's or Bob's cheating probability, Protocol \ref{Def: Classical rs protocol} then achieves a lower cheating probability for the other party.
Therefore, it is sufficient to compare the quantum Rabin OT protocol using two pure states to Protocol \ref{Def: Classical rs protocol} only, since the quantum protocol needs to outperform this classical protocol to exhibit a quantum advantage.

\section{Generalisation of the classical protocol with probabilistic choices by Alice and Bob} \label{App: Generalisation of random factor protocol}

A generalisation of Protocol \ref{Def: Classical rs protocol} is a protocol where Alice and Bob probabilistically choose between different versions of Protocol \ref{Def: Classical rs protocol}. That is, the most general approach is to probabilistically choose between $k$ different protocols of type \ref{Def: Classical rs protocol} with their own probabilities $s_k$ for Alice to send a bit and probabilities $r_k$ for Bob to read the bit value. Each of the $k$ different sub-protocols has a respective probability $p_k$ to occur, where $\sum_k p_k = 1$, and their own respective cheating probabilities for Alice and Bob, given by
\begin{align}
A^{k}_{OT} &=
\begin{cases}
1-r_k s_k = p^k_? &\text{ for } r_k < 1/2 \\
1-s_k +r_k s_k = 2 - p^k_? -s_k &\text{ for } r_k \geq 1/2, \end{cases} \nonumber \\
B^{k}_{OT} &= \frac{1}{2} (1+s_k) .
\end{align}

As explained in more detail in Section \ref{sec: Comparison Classical}, Bob's cheating probability for fixed values of $p^k_{?}$ and of Alice's cheating probability can be made higher in the case when $r_k < 1/2$ than in the case when $r_k \geq 1/2$. This occurs because of different combinations of values for the parameters $r_k$ and $s_k$, which result in the same value for $p^k_?$. Alice and Bob have to agree on values for these two parameters and because their objectives are going to be to lower the cheating probability of the respective other party, Alice will not accept any of the protocols where $r_k < 1/2$ because these favour Bob's cheating probability.

The probability of an honest Bob to not receive the bit in the general protocol is $p_? = 1 - \sum_k p_k r_k s_k$. Since Alice's and Bob's best cheating strategies are to cheat in each of the sub-protocols individually, we have
\begin{equation}
A^{\text{gen}}_{OT} = \sum_k p_k A^{k}_{OT} \quad \text{ and } \quad B^{\text{gen}}_{OT} = \sum_k p_k B^{k}_{OT} .
\end{equation}

Substituting in the expressions for $A^{k}_{OT}$ (for $r_k \geq 1/2$) and $B^{k}_{OT}$ yields
\begin{align}
A^{\text{gen}}_{OT} &= 1 - \sum_k p_k s_k + \sum_k p_k r_k s_k , \nonumber \\
B^{\text{gen}}_{OT} &= \dfrac{1}{2} \Big(1 + \sum_k p_k s_k \Big) . 
\end{align}
Using $\sum_k p_k s_k = 2 B^{\text{gen}}_{OT} - 1$ and $p_? = 1 - \sum_k p_k r_k s_k$, we obtain a tradeoff relation between Alice's and Bob's cheating probabilities,
\begin{equation}
B^{\text{gen}}_{OT} = \dfrac{1}{2} \big(3 - p_? - A^{\text{gen}}_{OT} \big) .
\end{equation}

This is the same tradeoff relation as in Eq. \eqref{eq: Tradeoff random factor} for the separate Protocol \ref{Def: Classical rs protocol}. We can conclude that the protocol with one set of probabilities $r$ and $s$ and its generalisation perform equally well. It is therefore sufficient to focus on Protocol \ref{Def: Classical rs protocol} only in the comparison of the Rabin oblivious transfer protocols in Section \ref{sec: Comparison}.

\section{Classical Rabin OT protocol expressed as quantum Rabin OT protocol using mixed states} \label{App: Mixed States Protocol}

The classical Rabin OT protocol, Protocol \ref{Def: Classical rs protocol}, where Alice sends a bit with some probability $s$ and Bob chooses to read the bit value with some probability $r$, can be expressed as a quantum protocol where Alice sends one of two mixed states, and Bob performs a quantum measurement on the state he receives.
Note that this quantum protocol can obviously be implemented using only classical means. To show that protocols with mixed states can in some cases lead to lower cheating probabilities than protocols with pure states, it is worthwhile to illustrate how the classical protocol, which for some $p_?$ outperforms the quantum protocol with pure states, can be expressed as a quantum protocol with mixed states.
The quantum protocol based on mixed states is described as follows.

\begin{enumerate}
\item Alice randomly chooses a bit value $x$, with equal probability for $x=0$ and $x=1$. She prepares and sends the state $\rho_x$ to Bob, where ($0 \leq s \leq 1$)
\begin{align}
\rho_0 = (1-s) \ket{0} \bra{0} + s \ket{1} \bra{1} , \nonumber \\
\rho_1 = (1-s) \ket{0} \bra{0} + s \ket{2} \bra{2} .
\end{align}
\item Bob performs an unambiguous discrimination measurement~\cite{Ivanovic87, Dieks88, Peres88} on the received state, where his measurement operators are ($0 \leq r \leq 1$)
\begin{align}
\Pi_? &= \ket{0} \bra{0} + (1-r) \ket{1} \bra{1} + (1-r) \ket{2} \bra{2}, \nonumber \\
\Pi_0 &= r \ket{1} \bra{1} , \nonumber \\
\Pi_1 &= r \ket{2}\bra{2}.
\end{align}
Bob will fail to learn the bit value with probability $p_? = 1 - rs$.
\end{enumerate}

When Alice is dishonest, she can cheat perfectly by sending $\ket{0}$, the state orthogonal to Bob's measurement operators $\Pi_0$ and $\Pi_1$, and she will know that Bob will never receive a bit. Thus, Bob needs to carry out some testing. When the protocol is repeated many times, by monitoring the probabilities of obtaining a bit or not, and checking if, on average, these probabilities are what he expects them to be, Bob can force a dishonest Alice to send the correct statistical mixture of the states $\ket{0}$, $\ket{1}$, and $\ket{2}$. Since Bob expects to receive a bit with probability $p_{\text{bit}} = rs$ and each of the bit values with equal probability, Alice, when cheating, needs to send $\ket{1}$ and $\ket{2}$ with a probability of $s/2$ each and $\ket{0}$ with probability $(1-s)$. Depending on the values of $s$ and $r$ and the knowledge of what state she sent to Bob, Alice will guess the most likely outcome on Bob's side. Her cheating probability is therefore given by
\begin{equation} \label{eq: Alice mixed state cheating}
A^{mq}_{OT} =
\begin{cases}
1-rs = p_? &\text{ for } r < 1/2 \\
1-s+rs = 2 - p_? -s &\text{ for } r \geq 1/2 . \end{cases}
\end{equation}

When Bob is dishonest, he applies a minimum-error measurement to distinguish between $\rho_0$ and $\rho_1$~\cite{Helstrom76} and his cheating probability is given by
\begin{equation} \label{eq: Bob mixed state cheating}
B^{mq}_{OT} = \frac{1}{2} (1 + s) .
\end{equation}

This quantum protocol using mixed states is equivalent to the classical protocol, Protocol \ref{Def: Classical rs protocol}. In other words, quantum protocols which use mixed states can sometimes outperform quantum protocols which use pure states.

\bibliography{Pure_State_Rabin_OT}

\begin{thebibliography}{29}%
\makeatletter
\providecommand \@ifxundefined [1]{%
 \@ifx{#1\undefined}
}%
\providecommand \@ifnum [1]{%
 \ifnum #1\expandafter \@firstoftwo
 \else \expandafter \@secondoftwo
 \fi
}%
\providecommand \@ifx [1]{%
 \ifx #1\expandafter \@firstoftwo
 \else \expandafter \@secondoftwo
 \fi
}%
\providecommand \natexlab [1]{#1}%
\providecommand \enquote  [1]{``#1''}%
\providecommand \bibnamefont  [1]{#1}%
\providecommand \bibfnamefont [1]{#1}%
\providecommand \citenamefont [1]{#1}%
\providecommand \href@noop [0]{\@secondoftwo}%
\providecommand \href [0]{\begingroup \@sanitize@url \@href}%
\providecommand \@href[1]{\@@startlink{#1}\@@href}%
\providecommand \@@href[1]{\endgroup#1\@@endlink}%
\providecommand \@sanitize@url [0]{\catcode `\\12\catcode `\$12\catcode
  `\&12\catcode `\#12\catcode `\^12\catcode `\_12\catcode `\%12\relax}%
\providecommand \@@startlink[1]{}%
\providecommand \@@endlink[0]{}%
\providecommand \url  [0]{\begingroup\@sanitize@url \@url }%
\providecommand \@url [1]{\endgroup\@href {#1}{\urlprefix }}%
\providecommand \urlprefix  [0]{URL }%
\providecommand \Eprint [0]{\href }%
\providecommand \doibase [0]{http://dx.doi.org/}%
\providecommand \selectlanguage [0]{\@gobble}%
\providecommand \bibinfo  [0]{\@secondoftwo}%
\providecommand \bibfield  [0]{\@secondoftwo}%
\providecommand \translation [1]{[#1]}%
\providecommand \BibitemOpen [0]{}%
\providecommand \bibitemStop [0]{}%
\providecommand \bibitemNoStop [0]{.\EOS\space}%
\providecommand \EOS [0]{\spacefactor3000\relax}%
\providecommand \BibitemShut  [1]{\csname bibitem#1\endcsname}%
\let\auto@bib@innerbib\@empty
\bibitem [{\citenamefont {Wiesner}(1983)}]{Wiesner83}%
  \BibitemOpen
  \bibfield  {author} {\bibinfo {author} {\bibfnamefont {S.}~\bibnamefont
  {Wiesner}},\ }\bibfield  {title} {\enquote {\bibinfo {title} {Conjugate
  coding},}\ }\href {\doibase 10.1145/1008908.1008920} {\bibfield  {journal}
  {\bibinfo  {journal} {SIGACT News}\ }\textbf {\bibinfo {volume} {15}},\
  \bibinfo {pages} {78} (\bibinfo {year} {1983})}\BibitemShut {NoStop}%
\bibitem [{\citenamefont {Even}\ \emph {et~al.}(1985)\citenamefont {Even},
  \citenamefont {Goldreich},\ and\ \citenamefont {Lempel}}]{Even85}%
  \BibitemOpen
  \bibfield  {author} {\bibinfo {author} {\bibfnamefont {S.}~\bibnamefont
  {Even}}, \bibinfo {author} {\bibfnamefont {O.}~\bibnamefont {Goldreich}}, \
  and\ \bibinfo {author} {\bibfnamefont {A.}~\bibnamefont {Lempel}},\
  }\bibfield  {title} {\enquote {\bibinfo {title} {A randomized protocol for
  signing contracts},}\ }\href {\doibase 10.1145/3812.3818} {\bibfield
  {journal} {\bibinfo  {journal} {Commun. ACM}\ }\textbf {\bibinfo {volume}
  {28}},\ \bibinfo {pages} {637} (\bibinfo {year} {1985})}\BibitemShut
  {NoStop}%
\bibitem [{\citenamefont {Brassard}\ \emph {et~al.}(1986)\citenamefont
  {Brassard}, \citenamefont {Cr{\'e}peau},\ and\ \citenamefont
  {Robert}}]{Brassard86}%
  \BibitemOpen
  \bibfield  {author} {\bibinfo {author} {\bibfnamefont {G.}~\bibnamefont
  {Brassard}}, \bibinfo {author} {\bibfnamefont {C.}~\bibnamefont
  {Cr{\'e}peau}}, \ and\ \bibinfo {author} {\bibfnamefont {J.-M.}\ \bibnamefont
  {Robert}},\ }\bibfield  {title} {\enquote {\bibinfo {title} {Information
  theoretic reductions among disclosure problems},}\ }in\ \href {\doibase
  10.1109/SFCS.1986.26} {\emph {\bibinfo {booktitle} {27th Annual Symposium on
  Foundations of Computer Science (sfcs 1986)}}}\ (\bibinfo  {publisher}
  {IEEE},\ \bibinfo {address} {Toronto, ON, Canada},\ \bibinfo {year} {1986})\
  p.\ \bibinfo {pages} {168}\BibitemShut {NoStop}%
\bibitem [{\citenamefont {Brassard}\ \emph {et~al.}(2003)\citenamefont
  {Brassard}, \citenamefont {Cr\'{e}peau},\ and\ \citenamefont
  {Wolf}}]{Brassard03}%
  \BibitemOpen
  \bibfield  {author} {\bibinfo {author} {\bibfnamefont {G.}~\bibnamefont
  {Brassard}}, \bibinfo {author} {\bibfnamefont {C.}~\bibnamefont
  {Cr\'{e}peau}}, \ and\ \bibinfo {author} {\bibfnamefont {S.}~\bibnamefont
  {Wolf}},\ }\bibfield  {title} {\enquote {\bibinfo {title} {Oblivious
  transfers and privacy amplification},}\ }\href {\doibase
  10.1007/s00145-002-0146-4} {\bibfield  {journal} {\bibinfo  {journal} {J.
  Cryptol.}\ }\textbf {\bibinfo {volume} {16}},\ \bibinfo {pages} {219}
  (\bibinfo {year} {2003})}\BibitemShut {NoStop}%
\bibitem [{\citenamefont {Rabin}(2005)}]{Rabin05}%
  \BibitemOpen
  \bibfield  {author} {\bibinfo {author} {\bibfnamefont {M.~O.}\ \bibnamefont
  {Rabin}},\ }\bibfield  {title} {\enquote {\bibinfo {title} {How to exchange
  secrets with oblivious transfer},}\ }\href@noop {} {\bibfield  {journal}
  {\bibinfo  {journal} {IACR Cryptol. ePrint Arch.}\ }\textbf {\bibinfo
  {volume} {2005}},\ \bibinfo {pages} {187} (\bibinfo {year}
  {2005})}\BibitemShut {NoStop}%
\bibitem [{\citenamefont {Cr{\'e}peau}(1988)}]{Crepeau88}%
  \BibitemOpen
  \bibfield  {author} {\bibinfo {author} {\bibfnamefont {C.}~\bibnamefont
  {Cr{\'e}peau}},\ }\bibfield  {title} {\enquote {\bibinfo {title} {Equivalence
  between two flavours of oblivious transfers},}\ }in\ \href@noop {} {\emph
  {\bibinfo {booktitle} {Advances in Cryptology — CRYPTO ’87}}}\ (\bibinfo
  {publisher} {Springer},\ \bibinfo {address} {Berlin, Heidelberg, Germany},\
  \bibinfo {year} {1988})\ p.\ \bibinfo {pages} {350}\BibitemShut {NoStop}%
\bibitem [{\citenamefont {Kilian}(1988)}]{Kilian88}%
  \BibitemOpen
  \bibfield  {author} {\bibinfo {author} {\bibfnamefont {J.}~\bibnamefont
  {Kilian}},\ }\bibfield  {title} {\enquote {\bibinfo {title} {Founding
  crytpography on oblivious transfer},}\ }in\ \href {\doibase
  10.1145/62212.62215} {\emph {\bibinfo {booktitle} {Proceedings of the
  Twentieth Annual ACM Symposium on Theory of Computing}}},\ \bibinfo {series
  and number} {STOC '88}\ (\bibinfo  {publisher} {ACM Press},\ \bibinfo
  {address} {New York, NY, USA},\ \bibinfo {year} {1988})\ p.~\bibinfo {pages}
  {20}\BibitemShut {NoStop}%
\bibitem [{\citenamefont {Mayers}(1997)}]{Mayers97}%
  \BibitemOpen
  \bibfield  {author} {\bibinfo {author} {\bibfnamefont {D.}~\bibnamefont
  {Mayers}},\ }\bibfield  {title} {\enquote {\bibinfo {title} {Unconditionally
  secure quantum bit commitment is impossible},}\ }\href {\doibase
  10.1103/PhysRevLett.78.3414} {\bibfield  {journal} {\bibinfo  {journal}
  {Phys. Rev. Lett.}\ }\textbf {\bibinfo {volume} {78}},\ \bibinfo {pages}
  {3414} (\bibinfo {year} {1997})}\BibitemShut {NoStop}%
\bibitem [{\citenamefont {Lo}(1997)}]{Lo97}%
  \BibitemOpen
  \bibfield  {author} {\bibinfo {author} {\bibfnamefont {H.-K.}\ \bibnamefont
  {Lo}},\ }\bibfield  {title} {\enquote {\bibinfo {title} {Insecurity of
  quantum secure computations},}\ }\href {\doibase 10.1103/PhysRevA.56.1154}
  {\bibfield  {journal} {\bibinfo  {journal} {Phys. Rev. A}\ }\textbf {\bibinfo
  {volume} {56}},\ \bibinfo {pages} {1154} (\bibinfo {year}
  {1997})}\BibitemShut {NoStop}%
\bibitem [{\citenamefont {Damg{\aa}rd}\ \emph {et~al.}(2008)\citenamefont
  {Damg{\aa}rd}, \citenamefont {Fehr}, \citenamefont {Salvail},\ and\
  \citenamefont {Schaffner}}]{Damgard08}%
  \BibitemOpen
  \bibfield  {author} {\bibinfo {author} {\bibfnamefont {I.~B.}\ \bibnamefont
  {Damg{\aa}rd}}, \bibinfo {author} {\bibfnamefont {S.}~\bibnamefont {Fehr}},
  \bibinfo {author} {\bibfnamefont {L.}~\bibnamefont {Salvail}}, \ and\
  \bibinfo {author} {\bibfnamefont {C.}~\bibnamefont {Schaffner}},\ }\bibfield
  {title} {\enquote {\bibinfo {title} {Cryptography in the bounded
  quantum-storage model},}\ }\href@noop {} {\bibfield  {journal} {\bibinfo
  {journal} {SIAM J. Comput.}\ }\textbf {\bibinfo {volume} {37}},\ \bibinfo
  {pages} {1865} (\bibinfo {year} {2008})}\BibitemShut {NoStop}%
\bibitem [{\citenamefont {Pital\'ua-Garc\'{\i}a}(2016)}]{Garcia16}%
  \BibitemOpen
  \bibfield  {author} {\bibinfo {author} {\bibfnamefont {D.}~\bibnamefont
  {Pital\'ua-Garc\'{\i}a}},\ }\bibfield  {title} {\enquote {\bibinfo {title}
  {Spacetime-constrained oblivious transfer},}\ }\href {\doibase
  10.1103/PhysRevA.93.062346} {\bibfield  {journal} {\bibinfo  {journal} {Phys.
  Rev. A}\ }\textbf {\bibinfo {volume} {93}},\ \bibinfo {pages} {062346}
  (\bibinfo {year} {2016})}\BibitemShut {NoStop}%
\bibitem [{\citenamefont {Pital\'ua-Garc\'{\i}a}\ and\ \citenamefont
  {Kerenidis}(2018)}]{Garcia18}%
  \BibitemOpen
  \bibfield  {author} {\bibinfo {author} {\bibfnamefont {D.}~\bibnamefont
  {Pital\'ua-Garc\'{\i}a}}\ and\ \bibinfo {author} {\bibfnamefont
  {I.}~\bibnamefont {Kerenidis}},\ }\bibfield  {title} {\enquote {\bibinfo
  {title} {Practical and unconditionally secure spacetime-constrained oblivious
  transfer},}\ }\href {\doibase 10.1103/PhysRevA.98.032327} {\bibfield
  {journal} {\bibinfo  {journal} {Phys. Rev. A}\ }\textbf {\bibinfo {volume}
  {98}},\ \bibinfo {pages} {032327} (\bibinfo {year} {2018})}\BibitemShut
  {NoStop}%
\bibitem [{\citenamefont {Pital\'ua-Garc\'{\i}a}(2019)}]{Garcia19}%
  \BibitemOpen
  \bibfield  {author} {\bibinfo {author} {\bibfnamefont {D.}~\bibnamefont
  {Pital\'ua-Garc\'{\i}a}},\ }\bibfield  {title} {\enquote {\bibinfo {title}
  {One-out-of-$m$ spacetime-constrained oblivious transfer},}\ }\href {\doibase
  10.1103/PhysRevA.100.012302} {\bibfield  {journal} {\bibinfo  {journal}
  {Phys. Rev. A}\ }\textbf {\bibinfo {volume} {100}},\ \bibinfo {pages}
  {012302} (\bibinfo {year} {2019})}\BibitemShut {NoStop}%
\bibitem [{\citenamefont {Chailloux}\ \emph {et~al.}(2013)\citenamefont
  {Chailloux}, \citenamefont {Kerenidis},\ and\ \citenamefont
  {Sikora}}]{Chailloux13}%
  \BibitemOpen
  \bibfield  {author} {\bibinfo {author} {\bibfnamefont {A.}~\bibnamefont
  {Chailloux}}, \bibinfo {author} {\bibfnamefont {I.}~\bibnamefont
  {Kerenidis}}, \ and\ \bibinfo {author} {\bibfnamefont {J.}~\bibnamefont
  {Sikora}},\ }\bibfield  {title} {\enquote {\bibinfo {title} {Lower bounds for
  quantum oblivious transfer},}\ }\href@noop {} {\bibfield  {journal} {\bibinfo
   {journal} {Quant. Inf. Comput.}\ }\textbf {\bibinfo {volume} {13}},\
  \bibinfo {pages} {158} (\bibinfo {year} {2013})}\BibitemShut {NoStop}%
\bibitem [{\citenamefont {Chailloux}\ \emph {et~al.}(2016)\citenamefont
  {Chailloux}, \citenamefont {Gutoski},\ and\ \citenamefont
  {Sikora}}]{Chailloux16}%
  \BibitemOpen
  \bibfield  {author} {\bibinfo {author} {\bibfnamefont {A.}~\bibnamefont
  {Chailloux}}, \bibinfo {author} {\bibfnamefont {G.}~\bibnamefont {Gutoski}},
  \ and\ \bibinfo {author} {\bibfnamefont {J.}~\bibnamefont {Sikora}},\
  }\bibfield  {title} {\enquote {\bibinfo {title} {Optimal bounds for
  semi-honest quantum oblivious transfer},}\ }\href@noop {} {\bibfield
  {journal} {\bibinfo  {journal} {Chicago J. Theor. Comput. Sci.}\ }\textbf
  {\bibinfo {volume} {2016}} (\bibinfo {year} {2016})}\BibitemShut {NoStop}%
\bibitem [{\citenamefont {Amiri}\ \emph {et~al.}(2021)\citenamefont {Amiri},
  \citenamefont {St\'arek}, \citenamefont {Reichmuth}, \citenamefont {Puthoor},
  \citenamefont {Mi\ifmmode~\check{c}\else \v{c}\fi{}uda}, \citenamefont
  {Mi\ifmmode~\check{s}\else \v{s}\fi{}ta}, \citenamefont
  {Du\ifmmode~\check{s}\else \v{s}\fi{}ek}, \citenamefont {Wallden},\ and\
  \citenamefont {Andersson}}]{Amiri21}%
  \BibitemOpen
  \bibfield  {author} {\bibinfo {author} {\bibfnamefont {R.}~\bibnamefont
  {Amiri}}, \bibinfo {author} {\bibfnamefont {R.}~\bibnamefont {St\'arek}},
  \bibinfo {author} {\bibfnamefont {D.}~\bibnamefont {Reichmuth}}, \bibinfo
  {author} {\bibfnamefont {I.~V.}\ \bibnamefont {Puthoor}}, \bibinfo {author}
  {\bibfnamefont {M.}~\bibnamefont {Mi\ifmmode~\check{c}\else \v{c}\fi{}uda}},
  \bibinfo {author} {\bibfnamefont {L.}~\bibnamefont {Mi\ifmmode~\check{s}\else
  \v{s}\fi{}ta}, \bibfnamefont {Jr.}}, \bibinfo {author} {\bibfnamefont
  {M.}~\bibnamefont {Du\ifmmode~\check{s}\else \v{s}\fi{}ek}}, \bibinfo
  {author} {\bibfnamefont {P.}~\bibnamefont {Wallden}}, \ and\ \bibinfo
  {author} {\bibfnamefont {E.}~\bibnamefont {Andersson}},\ }\bibfield  {title}
  {\enquote {\bibinfo {title} {Imperfect 1-out-of-2 quantum oblivious transfer:
  Bounds, a protocol, and its experimental implementation},}\ }\href {\doibase
  10.1103/PRXQuantum.2.010335} {\bibfield  {journal} {\bibinfo  {journal} {PRX
  Quantum}\ }\textbf {\bibinfo {volume} {2}},\ \bibinfo {pages} {010335}
  (\bibinfo {year} {2021})}\BibitemShut {NoStop}%
\bibitem [{\citenamefont {Stroh}\ \emph {et~al.}(2023)\citenamefont {Stroh},
  \citenamefont {Horov\'a}, \citenamefont {St\'arek}, \citenamefont {Puthoor},
  \citenamefont {Mi\ifmmode~\check{c}\else \v{c}\fi{}uda}, \citenamefont
  {Du\ifmmode~\check{s}\else \v{s}\fi{}ek},\ and\ \citenamefont
  {Andersson}}]{Stroh23}%
  \BibitemOpen
  \bibfield  {author} {\bibinfo {author} {\bibfnamefont {L.}~\bibnamefont
  {Stroh}}, \bibinfo {author} {\bibfnamefont {N.}~\bibnamefont {Horov\'a}},
  \bibinfo {author} {\bibfnamefont {R.}~\bibnamefont {St\'arek}}, \bibinfo
  {author} {\bibfnamefont {I.~V.}\ \bibnamefont {Puthoor}}, \bibinfo {author}
  {\bibfnamefont {M.}~\bibnamefont {Mi\ifmmode~\check{c}\else \v{c}\fi{}uda}},
  \bibinfo {author} {\bibfnamefont {M.}~\bibnamefont {Du\ifmmode~\check{s}\else
  \v{s}\fi{}ek}}, \ and\ \bibinfo {author} {\bibfnamefont {E.}~\bibnamefont
  {Andersson}},\ }\bibfield  {title} {\enquote {\bibinfo {title}
  {Noninteractive {XOR} quantum oblivious transfer: Optimal protocols and their
  experimental implementations},}\ }\href {\doibase
  10.1103/PRXQuantum.4.020320} {\bibfield  {journal} {\bibinfo  {journal} {PRX
  Quantum}\ }\textbf {\bibinfo {volume} {4}},\ \bibinfo {pages} {020320}
  (\bibinfo {year} {2023})}\BibitemShut {NoStop}%
\bibitem [{\citenamefont {Bansal}\ and\ \citenamefont
  {Sikora}(2023)}]{Bansal23}%
  \BibitemOpen
  \bibfield  {author} {\bibinfo {author} {\bibfnamefont {A.}~\bibnamefont
  {Bansal}}\ and\ \bibinfo {author} {\bibfnamefont {J.}~\bibnamefont
  {Sikora}},\ }\href@noop {} {\enquote {\bibinfo {title} {Breaking barriers in
  two-party quantum cryptography via stochastic semidefinite programming},}\ }
  (\bibinfo {year} {2023}),\ \Eprint {http://arxiv.org/abs/2304.13200}
  {arXiv:2304.13200 [quant-ph]} \BibitemShut {NoStop}%
\bibitem [{\citenamefont {He}\ and\ \citenamefont {Wang}(2006)}]{He06}%
  \BibitemOpen
  \bibfield  {author} {\bibinfo {author} {\bibfnamefont {G.~P.}\ \bibnamefont
  {He}}\ and\ \bibinfo {author} {\bibfnamefont {Z.~D.}\ \bibnamefont {Wang}},\
  }\bibfield  {title} {\enquote {\bibinfo {title} {Oblivious transfer using
  quantum entanglement},}\ }\href {\doibase 10.1103/PhysRevA.73.012331}
  {\bibfield  {journal} {\bibinfo  {journal} {Phys. Rev. A}\ }\textbf {\bibinfo
  {volume} {73}},\ \bibinfo {pages} {012331} (\bibinfo {year}
  {2006})}\BibitemShut {NoStop}%
\bibitem [{\citenamefont {Peat}\ and\ \citenamefont
  {Andersson}(2024)}]{Peat24}%
  \BibitemOpen
  \bibfield  {author} {\bibinfo {author} {\bibfnamefont {J.~T.}\ \bibnamefont
  {Peat}}\ and\ \bibinfo {author} {\bibfnamefont {E.}~\bibnamefont
  {Andersson}},\ }\href@noop {} {\enquote {\bibinfo {title} {Cheating in
  quantum {R}abin oblivious transfer using delayed measurements},}\ } (\bibinfo
  {year} {2024}),\ \Eprint {http://arxiv.org/abs/2408.12388} {arXiv:2408.12388
  [quant-ph]} \BibitemShut {NoStop}%
\bibitem [{\citenamefont {Cheong}\ \emph {et~al.}(2010)\citenamefont {Cheong},
  \citenamefont {Hsieh},\ and\ \citenamefont {Koshiba}}]{Cheong10}%
  \BibitemOpen
  \bibfield  {author} {\bibinfo {author} {\bibfnamefont {K.~Y.}\ \bibnamefont
  {Cheong}}, \bibinfo {author} {\bibfnamefont {M.-H.}\ \bibnamefont {Hsieh}}, \
  and\ \bibinfo {author} {\bibfnamefont {T.}~\bibnamefont {Koshiba}},\
  }\href@noop {} {\enquote {\bibinfo {title} {A weak quantum oblivious
  transfer},}\ } (\bibinfo {year} {2010}),\ \Eprint
  {http://arxiv.org/abs/1010.0702} {arXiv:1010.0702 [quant-ph]} \BibitemShut
  {NoStop}%
\bibitem [{\citenamefont {Ivanovic}(1987)}]{Ivanovic87}%
  \BibitemOpen
  \bibfield  {author} {\bibinfo {author} {\bibfnamefont {I.~D.}\ \bibnamefont
  {Ivanovic}},\ }\bibfield  {title} {\enquote {\bibinfo {title} {How to
  differentiate between non-orthogonal states},}\ }\href {\doibase
  https://doi.org/10.1016/0375-9601(87)90222-2} {\bibfield  {journal} {\bibinfo
   {journal} {Phys. Lett. A}\ }\textbf {\bibinfo {volume} {123}},\ \bibinfo
  {pages} {257} (\bibinfo {year} {1987})}\BibitemShut {NoStop}%
\bibitem [{\citenamefont {Dieks}(1988)}]{Dieks88}%
  \BibitemOpen
  \bibfield  {author} {\bibinfo {author} {\bibfnamefont {D.}~\bibnamefont
  {Dieks}},\ }\bibfield  {title} {\enquote {\bibinfo {title} {Overlap and
  distinguishability of quantum states},}\ }\href {\doibase
  https://doi.org/10.1016/0375-9601(88)90840-7} {\bibfield  {journal} {\bibinfo
   {journal} {Phys. Lett. A}\ }\textbf {\bibinfo {volume} {126}},\ \bibinfo
  {pages} {303} (\bibinfo {year} {1988})}\BibitemShut {NoStop}%
\bibitem [{\citenamefont {Peres}(1988)}]{Peres88}%
  \BibitemOpen
  \bibfield  {author} {\bibinfo {author} {\bibfnamefont {A.}~\bibnamefont
  {Peres}},\ }\bibfield  {title} {\enquote {\bibinfo {title} {How to
  differentiate between non-orthogonal states},}\ }\href {\doibase
  https://doi.org/10.1016/0375-9601(88)91034-1} {\bibfield  {journal} {\bibinfo
   {journal} {Phys. Lett. A}\ }\textbf {\bibinfo {volume} {128}},\ \bibinfo
  {pages} {19} (\bibinfo {year} {1988})}\BibitemShut {NoStop}%
\bibitem [{\citenamefont {Helstrom}(1976)}]{Helstrom76}%
  \BibitemOpen
  \bibfield  {author} {\bibinfo {author} {\bibfnamefont {C.~W.}\ \bibnamefont
  {Helstrom}},\ }\href@noop {} {\emph {\bibinfo {title} {Quantum {D}etection
  and {E}stimation {T}heory}}}\ (\bibinfo  {publisher} {Academic Press},\
  \bibinfo {address} {New York},\ \bibinfo {year} {1976})\BibitemShut {NoStop}%
\bibitem [{\citenamefont {Hausladen}\ and\ \citenamefont
  {Wootters}(1994)}]{Hausladen94}%
  \BibitemOpen
  \bibfield  {author} {\bibinfo {author} {\bibfnamefont {P.}~\bibnamefont
  {Hausladen}}\ and\ \bibinfo {author} {\bibfnamefont {W.~K.}\ \bibnamefont
  {Wootters}},\ }\bibfield  {title} {\enquote {\bibinfo {title} {A ‘pretty
  good’ measurement for distinguishing quantum states},}\ }\href {\doibase
  10.1080/09500349414552221} {\bibfield  {journal} {\bibinfo  {journal} {J.
  Mod. Opt.}\ }\textbf {\bibinfo {volume} {41}},\ \bibinfo {pages} {2385}
  (\bibinfo {year} {1994})}\BibitemShut {NoStop}%
\bibitem [{\citenamefont {Andersson}\ \emph {et~al.}(2002)\citenamefont
  {Andersson}, \citenamefont {Barnett}, \citenamefont {Gilson},\ and\
  \citenamefont {Hunter}}]{Andersson02}%
  \BibitemOpen
  \bibfield  {author} {\bibinfo {author} {\bibfnamefont {E.}~\bibnamefont
  {Andersson}}, \bibinfo {author} {\bibfnamefont {S.~M.}\ \bibnamefont
  {Barnett}}, \bibinfo {author} {\bibfnamefont {C.~R.}\ \bibnamefont {Gilson}},
  \ and\ \bibinfo {author} {\bibfnamefont {K.}~\bibnamefont {Hunter}},\
  }\bibfield  {title} {\enquote {\bibinfo {title} {Minimum-error discrimination
  between three mirror-symmetric states},}\ }\href {\doibase
  10.1103/PhysRevA.65.052308} {\bibfield  {journal} {\bibinfo  {journal} {Phys.
  Rev. A}\ }\textbf {\bibinfo {volume} {65}},\ \bibinfo {pages} {052308}
  (\bibinfo {year} {2002})}\BibitemShut {NoStop}%
\bibitem [{\citenamefont {Damg{\aa}rd}\ \emph {et~al.}(2004)\citenamefont
  {Damg{\aa}rd}, \citenamefont {Fehr}, \citenamefont {Morozov},\ and\
  \citenamefont {Salvail}}]{Damgard04}%
  \BibitemOpen
  \bibfield  {author} {\bibinfo {author} {\bibfnamefont {I.~B.}\ \bibnamefont
  {Damg{\aa}rd}}, \bibinfo {author} {\bibfnamefont {S.}~\bibnamefont {Fehr}},
  \bibinfo {author} {\bibfnamefont {K.}~\bibnamefont {Morozov}}, \ and\
  \bibinfo {author} {\bibfnamefont {L.}~\bibnamefont {Salvail}},\ }\bibfield
  {title} {\enquote {\bibinfo {title} {Unfair noisy channels and oblivious
  transfer},}\ }in\ \href@noop {} {\emph {\bibinfo {booktitle} {Theory of
  Cryptography}}},\ \bibinfo {editor} {edited by\ \bibinfo {editor}
  {\bibfnamefont {M.}~\bibnamefont {Naor}}}\ (\bibinfo  {publisher}
  {Springer},\ \bibinfo {address} {Berlin},\ \bibinfo {year} {2004})\ p.\
  \bibinfo {pages} {355}\BibitemShut {NoStop}%
\bibitem [{\citenamefont {Mochon}(2007)}]{Mochon07}%
  \BibitemOpen
  \bibfield  {author} {\bibinfo {author} {\bibfnamefont {C.}~\bibnamefont
  {Mochon}},\ }\href@noop {} {\enquote {\bibinfo {title} {Quantum weak coin
  flipping with arbitrarily small bias},}\ } (\bibinfo {year} {2007}),\ \Eprint
  {http://arxiv.org/abs/0711.4114} {arXiv:0711.4114 [quant-ph]} \BibitemShut
  {NoStop}%
\end{thebibliography}%

\end{document}